\documentclass[reprint,nofootinbib,showpacs]{revtex4}
\usepackage{graphicx}  % needed for figures
\usepackage{dcolumn}   % needed for some tables
\usepackage{bm}        % for math
\usepackage{amssymb}
\usepackage{hyperref}
\usepackage{multirow}

\begin{document}
 \title{Implication of $b \to c\tau\nu$ flavor anomalies on $B_s\rightarrow {D_s^*}\tau\nu$ decay observables}

     \author{Nilakshi~Das${}$}
     \email{nilakshi@rs.phy.student.nits.ac.in} 
     \author{Rupak~Dutta${}$}
     \email{rupak@phy.nits.ac.in}
    \affiliation{
    National Institute of Technology Silchar, Silchar 788010, India\\
    }

\begin{abstract}
The experimental predictions of $R_D$, $R_{D^*}$, $R_{J/\psi}$, $P_\tau^{D^*}$ and $F_L^{D^*}$ in $B$ decays 
mediated via $b\,\to\, c\, l\,\nu$ quark level transition 
deviate significantly from the standard model expectations. The current world average of the ratio of branching ratios $R_D$ and $R_{D^*}$ 
in $B\,\to\, D^{(*)}\, l\,\nu$ ($l\in e,\tau$) show $1.4\sigma$ and $2.5\sigma$ deviation from the SM expectations. 
Similarly, the $\tau$ polarization fraction $P_{\tau}^{D^*}$ and the longitudinal polarization fraction of the $D^*$ meson $F_L^{D^*}$ 
in $B\,\to\, D^*\,\tau\,\nu$ are found to deviate from the standard model expectations at $1.6\sigma$ and $1.5\sigma$ level, respectively.
In addition, the ratio of branching ratio $R_{J/\psi}$ in $B_c\,\to\,{J/\psi}\, l\,\nu$ deviates from the standard model  prediction at 
$2\sigma$ level. In this regard, we study the implication of $R_D$, $R_{D^*}$, $R_{J/\psi}$, $P_\tau^{D^*}$, and $F_L^{D^*}$ anomalies on 
$B_s\,\to\, {D^*_s}\,\tau\,\nu$ 
decay observables in a model independent effective field theory formalism.
We give predictions of several physical observables in the standard model and in the presence of various $1D$ and $2D$ new physics scenarios.
\end{abstract}
 %     leptonic + semileptonic decays

\maketitle

\section{Introduction}
The standard model (SM) of particle physics, a self-consistent theory, encompasses most of the successful
evidences in understanding the fundamental particles and their interactions. 
Although it has provided a practical exhibition and explanation for most of the important experimental 
predictions till date, it, however, failed to explain various long standing phenomena such as
matter-antimatter asymmetry of the universe, the hierarchy problem, dark matter, neutrino mass etc.
Hence, it is necessary to look for physics that lies beyond the SM. 
The process of searching for new physics~(NP) can be performed in two ways: one is the direct search and other is the 
indirect search.
The direct searches include direct detection of new particles and their interactions at the ongoing or future colliders,
whereas, no such direct evidences have been reported at the experiments so far. 
On the other hand, the indirect searches are more concerned 
towards the possible indirect effects of these new particles on low energy processes. Among the various indirect searches,  
the flavor changing charged current (FCCC) and the flavor changing neutral current
(FCNC) quark level transitions at the electroweak scale which account for the lepton flavor universality 
violation (LFUV) in the $B$ sectors have been the ideal places to look for new physics effects.

The SM of particle physics assumes that the leptons couple to the gauge bosons with equal strength irrespective
of their generations. This condition of the lepton flavor universality (LFU) contradicts the various
experimental measurements in $b\,\to\, (c,\,u)\, l\,\nu$ and $b\, \to\, s\, l^+\, l^-$
flavor changing quark level transitions. To unravel the flavor structure of the $b$ flavor mesons at the 
electroweak scale various theoretically clean flavor ratios such as, 
$R_D$, $R_{D^*}$, $R_{J/\psi}$, $P_\tau^{D^*}$, and $F_L^{D^*}$ have been defined. Those are

\begin{eqnarray}
R_{D^{(*)}}=\frac{\mathcal{B}\,(B \,\to \, D^{(*)}\, \tau \, \nu)}{\mathcal{B}\,(B\,\to D^{(*)}\, \{e/\mu\}\,\nu)}; \hspace{1cm}
R_{J/\psi}=\frac{\mathcal{B}\,(B_c\, \to  J/\psi \, \tau \,\nu)}{\mathcal{B}\,(B_c \,\to \, J/\psi \, \{e/\mu\}\,\nu)}
\end{eqnarray}

\begin{eqnarray}
P_{\tau}^{D^*} = \frac{{d\Gamma^{D^*}(+)}/{dq^2} - {d\Gamma^{D^*}(-)}/{dq^2}}{{d\Gamma^{D^*}}/{dq^2}}; \hspace{1cm}
F_L^{D^*}=\frac{\Gamma\,(B\,\to\, D_L^*\,\tau\,\nu)}{\Gamma\,(B\,\to\, D^*\,\tau\,\nu)}
\end{eqnarray}

At present, we have the precise $B \to D$ form factors that have been calculated using the lattice quantum chromodynamics
(LQCD) technique from various groups. In fact these lattice calculations provide a very precise SM predictions
of the ratio of branching ratio $R_D\,=\,0.299 \pm 0.003$ in $B\,\to D\,l\,\nu$ decay mode~\cite{Lattice:2015rga, Na:2015kha,
Aoki:2016frl,Bigi:2016mdz}. As of $B \to D^*$ lattice QCD form factors are concerned, at present only some unquenched calculations 
at the zero recoil exists from the Fermilab Lattice and MILC Collaborations~\cite{Bernard:2008dn,Bailey:2014tva}. The non zero recoil 
calculations for 
the $B \to D^*$ form factors are limited by the availability of computational resources and the efficient algorithms.
Apart from that there are various different SM predictions of $R_{D^*}$ available in literature~\cite{Bernlochner:2017jka,Jaiswal:2017rve,
Fajfer:2012vx,Bigi:2017jbd}. The arithmetic average of $R_{D^*}$ is reported to be $R_{D^*}\,=\,0.258 \pm 0.005$ by the Heavy Flavor Averaging group.
The SM predictions of $R_{J/\psi}$ in $B_c \,\to \, J/\psi \, {l}\,\nu$ decay mode was obtained using various form factors~\cite{Ivanov:2000aj,Ebert:2003cn,AbdElHady:1999xh,Wen-Fei:2013uea,Hsiao:2016pml,Dutta:2017xmj,Dutta:2017wpq}
and recently in Ref.~\cite{Cohen:2018dgz}, the authors report the $R_{J/\psi}$ bound to be $[0.20, 0.39]$ at the $95\%$ confidence level.
Similar to $B \to D^*$ form factors, the precise $B_c \,\to \, J/\psi$ form factors from the lattice QCD are still awaited.
Another interesting observable is the lepton polarization fraction $P_\tau^{D^*}$ considered along the longitudinal direction of 
the $\tau$ lepton in $B\,\to\, D^*\, \tau\,\nu$ decay mode and it turns out to be very sensitive to the various NP models.
In SM, the $P_\tau^{D^*}$ is 
predicted to be $−0.497 \pm 0.013$~\cite{Tanaka:2012nw}, whereas, it is predicted to be in the range $[-0.6, 1.0]$ and 
$[-0.5, 0.0]$~\cite{Sakaki:2013bfa} in type - II $2$HDM and leptoquark model, respectively.
In addition to the $\tau$ polarization fraction, the longitudinal polarization fraction of the $D^*$ meson $F_L^{D^*}$ in 
$B\to D^*\, \tau \, \nu$ decay mode can, in principle, help in distinguishing between the new scalar and tensor NP
Lorentz structures. The SM predictions of this observable is reported to be $F_L^{D^*}=0.46 \pm 0.04$~\cite{Alok:2016qyh}. 

The experimental measurements of these observables differ significantly from the SM expectations.
There has been various measurements of $R_D$ and $R_{D^*}$ from BABAR, Belle and LHCb. The current world average of $R_D$ and
$R_{D^*}$ stands at $1.4\sigma$ and $2.5\sigma$ away from the SM prediction. The combined deviation of $R_D$-$R_{D^*}$ is reported to be 
$3.08\sigma$
away from the SM expectations. Similarly, LHCb measurement of $R_{J/\psi}$ in 2017~\cite{Aaij:2017tyk} stands at more than $2\sigma$ 
away from the standard model expectation. As the error band in $R_{J/\psi}$ measurement is very large, 
subsequent measurements may help in reducing the the large systematic uncertainty. 
The consecutive measurements of $P_\tau^{D^*}$ at Belle~\cite{Hirose:2016wfn, Hirose:2017dxl} also show $1.6\sigma$ deviation from
the SM expectations. Similarly, the preliminary result pertaining to the measurement of $F_L^{D^*}$ at Belle shows~\cite{Abdesselam:2019wbt} $1.5\sigma$
deviation from the SM expectations. It should be noted that future Belle II measurements of $P_\tau^{D^*}$ and 
$F_L^{D^*}$ with higher precision can provide better complimentary information regarding NP in $b \to c\tau\nu$ decays.  
For completeness we report the SM and the experimental results of all the observables in Table~\ref{smexpt}.

\begin{table}[htbp]
\centering 
\begin{tabular}{|l|l|l|}
\hline
      & Standard model prediction & Experimental prediction \\
\hline
\hline
$R_D$ & $0.299 \pm 0.003$~\cite{Lattice:2015rga, Na:2015kha,Aoki:2016frl,Bigi:2016mdz} & $0.340 \pm 0.027 \pm 0.013$~\cite{Lees:2012xj,Lees:2013uzd,Huschle:2015rga,Abdesselam:2019dgh} \\
\hline
$R_{D^*}$ & $0.258 \pm 0.005$~\cite{Bernlochner:2017jka,Jaiswal:2017rve,Fajfer:2012vx,Bigi:2017jbd} & $0.295 \pm 0.011 \pm 0.008$~\cite{Lees:2012xj,Lees:2013uzd,Huschle:2015rga,Abdesselam:2019dgh, Sato:2016svk, Hirose:2016wfn,Hirose:2017dxl,Aaij:2015yra, Aaij:2017uff,Aaij:2017deq} \\
\hline
$R_{J/\psi}$ & $[0.20, 0.39]$~\cite{Cohen:2018dgz} & $0.71 \pm 0.17 \pm 0.18$~\cite{Aaij:2017tyk} \\
\hline
$P_\tau^{D^*}$ & $ −0.497 \pm 0.013$~\cite{Tanaka:2012nw} & $-0.38 \pm 0.51^{+0.21}_{-0.16}$~\cite{Hirose:2016wfn, Hirose:2017dxl} \\
\hline
$F_L^{D^*}$ & $0.46 \pm 0.04$~\cite{Alok:2016qyh}   & $0.60 \pm 0.08 \pm 0.035$~\cite{Abdesselam:2019wbt}\\
\hline
\hline
\end{tabular}
\caption{Current status of $R_{D^{(*)}}$, $R_{J/\psi}$, $P_\tau^{D^*}$ and $F_L^{D^*}$}
\label{smexpt}
\end{table}

It was shown in Ref.~\cite{Alonso:2016oyd} that, the lifetime of $B_c$ meson have a serious impact on scalar NP Lorentz structures. 
The SM prediction of the lifetime of $B_c$ meson demands that the 
fraction of the branching ratio of $\mathcal{B}(B_c\to \tau\nu)$ cannot exceed the total width. This has put a severe
constraint on scalar NP couplings. 
In SM, the lifetime of $B_c$ meson $\tau_{B_c}=0.52_{-0.12}^{+0.18}$ ps~\cite{Chang:2000ac} is obtained 
by using operator product expansion
and in fact it is consistent with the experimental value of $\tau_{B_c}=0.507(9)$ ps~\cite{Tanabashi:2018oca}. 
Although the $B_c \to \tau\nu$ branching ratio should be less than or equal to $5\%$, this constraint can be relaxed upto $30\%$ if
the upper bound of $\tau_{B_c}$ is considered~\cite{Alonso:2016oyd}.
Moreover, recent LEP data taken at the Z peak requires 
the branching ratio of $\mathcal{B}(B_c\to \tau\nu)$ to be less than or equal to $10\%$. This is significantly a stronger constraint
compared to the $\mathcal{B}(B_c\to \tau\nu)\le 30\%$ obtained from the lifetime of $B_c$ meson~\cite{Akeroyd:2017mhr}. On the other hand
by considering all the possible uncertainties, a highly relaxed bound of $60\%$ is also allowed for the $B_c \to \tau\nu$ branching
fraction.
A comparison among the three different bounds of $10\%$, $30\%$ and $60\%$ have been well studied in Ref.~\cite{Blanke:2019qrx}.
Nevertheless, in this paper we consider the stronger bound of $\mathcal{B}(B_c\to \tau\nu)\le 10\%$ for our NP analysis.

In order to explain these anomalies, various model dependent and model independent analysis have been performed.
An incomplete list of literature can be found in the Refs.
~\cite{Sakaki:2014sea,Freytsis:2015qca,
Bhattacharya:2016zcw,Alok:2017qsi,Azatov:2018knx,Bifani:2018zmi, Huang:2018nnq,Hu:2018veh,Feruglio:2018fxo,Jung:2018lfu,Datta:2017aue,
Bernlochner:2018kxh,Alok:2018uft,Fajfer:2012jt,Crivellin:2012ye,Li:2016vvp,Bhattacharya:2016mcc,Leljak:2019fqa,Becirevic:2019tpx,
Dutta:2015ueb,Dutta:2016eml,Dutta:2018zqp,Dutta:2018jxz,Rajeev:2018txm,Dutta:2018vgu,Rajeev:2019ktp,Dutta:2019wxo,Bardhan:2016uhr,Gomez:2019xfw,Alok:2019uqc,Yan:2019hpm} 
To this end, the long standing anomalies persisting in the flavor sector motivate us to study the $B_s\,\to\, D^{*}_{s}\,l\, \nu$ decay mode which undergo 
similar $b\, \to\, c\,$ quark level transition.
The study corresponding to $B_s\,\to\, D^{*}_{s}\,l\, \nu$ decay mode is of great interest in the present experiments as
this decay mode will serve as an important channel since,
both $B\, \to\, D^*\, l\, \nu$ and $B_s\,\to\, D^{*}_{s}\,l\,\nu$ decay modes undergo similar $b\,\to\, c$ quark level 
transitions and under the SU(3) flavor symmetry both the decay modes exhibit similar properties.

The $B_s\,\to\, D^{*}_{s}\,l\,\nu$ decay mode has been studied by various authors in SM with 
different form factors obtained using the constituent quark meson (CQM) model~\cite{Zhao:2006at},
the QCD sum rule~\cite{Azizi:2008vt, Bayar:2008cv}, the light cone sum rule (LCSR)~\cite{Li:2009wq},
the covariant light-front quark model (CLFQM)~\cite{Li:2010bb}, the instantaneous Bethe-Salpeter
equation~\cite{Chen:2011ut,Zhou:2019stx} and lattice QCD at zero recoil point~\cite{Harrison:2017fmw}. 
In Ref.~\cite{Fan:2013kqa}, the author predicts the SM expectation of $R_{D_s^*}$ to be $0.302 \pm 0.011$. 
In Ref.~\cite{Cohen:2019zev}, the authors give predictions of $R_{D_s^*}$, $P_{\tau}^{D_s^*}$,
and $F_L^{D_s^*}$ by considering the BGL parametrization of lattice QCD data.
Very recently in Ref.~\cite{Sahoo:2019hbu}, the authors perform a model independent analysis on 
$B_s\,\to\, D_s^*\,l\,\nu$ decay mode by considering pQCD form factors. Also in Ref~\cite{Hu:2019bdf}, the authors discuss the SM results of $R_{D_s^*}$, $P_{\tau}^{D_s^*}$, $F_L^{D_s^*}$ and $A_{FB}^{\tau}$ in $B_s\,\to\, D_s^*\,l\,\nu$ decay mode by using the form factors obtained by employing the  pQCD factorization formalism combining with the lattice QCD inputs.
In the present paper, we follow a model independent effective field theory formalism and study the implications of $R_D, R_{D^*}$, $R_{J/\psi}$, $P_\tau^{D^*}$ and $F_L^{D^*}$ on $B_s\,\to D^{*}_{s}\, \tau\,\nu$ decay mode. 
We give predictions of various physical observables such as the branching ratio, the ratio 
of branching ratio, the forward backward asymmetry, the longitudinal polarization fraction of the charged lepton, the convexity parameter, the forward backward asymmetry of transversely polarized $D^*_{s}$ meson, 
and the longitudinal polarization fraction of the $D^*_{s}$ meson 
within the SM and in the presence of various NP couplings. 
In our analysis we consider an indirect constrain coming from $\mathcal{B}(B_c\to \tau\nu) \le 10\%$~\cite{Akeroyd:2017mhr}.
Our analysis significantly differs from~\cite{Sahoo:2019hbu} for several reasons:
First, we use the form factors calculated in relativistic quark model and we
consider the constraints coming from $P_\tau^{D^*}$ and $F_L^{D^*}$ in addition to $R_D, R_{D^*}$, $R_{J/\psi}$. 
Second, we have considered the effects coming from the right handed neutrino couplings in addition to the 
left handed neutrino couplings.
Third, the results pertaining to the convexity parameter, the forward backward asymmetry of transversely polarized 
$D^*_{s}$ meson and the longitudinal polarization fraction of the $D^*_{s}$ meson have been discussed
in addition to the various other observables.

The paper is organized as follows. In section~\ref{ph}, we start with the most general effective Lagrangian for $b\to c l \nu$
quark level transition in the presence of NP at the renormalization scale $\mu = m_b$. We report all the relevant
formulae such as the branching ratio, the ratio of branching ratio, the forward backward asymmetry,
the longitudinal polarization fraction of the charged lepton, the convexity parameter, 
the forward backward asymmetry of transversely polarized $D^*_{s}$ meson, and the longitudinal
polarization fraction of the $D^*_{s}$ meson. In section \ref{results}, we report our results within the SM and within various $1D$ and
$2D$ NP scenarios. We conclude with a brief summary of our results in 
section \ref{Conclusion}.  
      
\section{Phenomenology}
\label{ph}
The most general effective Lagrangian for $b\rightarrow cl\nu $ quark level transition in
the presence of new vector, scalar and tensor NP couplings can be written as~\cite{Bhattacharya:2011qm, Cirigliano:2009wk} 
\begin{eqnarray}
 \mathcal{L}_{eff}&=& - \frac{4G_F}{\sqrt{2}} |V_{cb}| \Bigg[ \left( 1 + {V_L} \right) \mathcal{O}_{V_L} + {V_R} \mathcal{O}_{V_R} + \widetilde{V}_L \widetilde{\mathcal{O}}_{V_L} + \widetilde{V}_R \widetilde{\mathcal{O}}_{V_R} +\nonumber\\
 &&
{S_L} \mathcal{O}_{S_L} + {S_R} \mathcal{O}_{S_R}+\widetilde{{S}}_L \widetilde{\mathcal{O}}_{S_L}  + \widetilde{{S}}_R \widetilde{\mathcal{O}}_{S_R} + {T}_L \mathcal{O}_{T}
+ \widetilde{{T}}_L \widetilde{\mathcal{O}}_{T_L}\Bigg] +{\rm h.c.} 
\end{eqnarray}
where,
\begin{eqnarray}
\mathcal{O}_{V_L} = \bar{l}_L \gamma_{\mu}\nu_L\bar{c}_L\gamma^\mu b_L ; \hspace{0.5cm}
\mathcal{O}_{V_R} = {l}_L\gamma_{\mu}\nu_L\bar{c}_R\gamma^{\mu}b_R; \hspace{0.5cm}
\mathcal{O}_{S_L} = \bar{l}_R\nu_l\bar{c}_Rb_L; \hspace{0.5cm}
\mathcal{O}_{S_R} = \bar{l}_R\nu_L\bar{c}_Lb_R;\hspace{0.5cm}\nonumber 
\widetilde{\mathcal{O}}_{V_L} = \bar{l}_R\gamma_{\mu}\nu_R\bar{c}_L\gamma^{\mu}b_L; \hspace{0.5cm}\\ 
\widetilde{\mathcal{O}}_{V_R} = \bar{l}_R\gamma_{\mu}\nu_R\bar{c}_R\gamma^{\mu}b_R; \hspace{0.5cm} 
\widetilde{\mathcal{O}}_{S_L} = \bar{l_L}\nu_R\bar{c}_Rb_L; \hspace{0.5cm}
\widetilde{\mathcal{O}}_{S_R} =  \bar{l}_L\nu_R\bar{c}_Lb_R; \hspace{0.5cm}
\mathcal{O}_{T_L} = \bar{l}_R\sigma_{\mu\nu}\nu_L\bar{c}_R\sigma^{\mu\nu}b_L; \hspace{0.3cm} 
\widetilde{\mathcal{O}}_{T_L} = \bar{l}_R\sigma_{\mu\nu}\nu_R\bar{c}_L\sigma^{\mu\nu}b_R \hspace{0.5cm}\nonumber
\end{eqnarray}
Here, $G_F$ and $|V_{cb}|$ represent the Fermi coupling constant and the Cabbibo-Kobayashi-Mashkawa(CKM) matrix element.
$V_{L,R}, S_{L,R},T_L$ are the new physics (NP) Wilson coefficient(WC) which involve left handed neutrinos whereas,
$\widetilde{V}_{L,R},\widetilde {S}_{L,R}, \widetilde{T}_L$ represent the WC's which involve the right handed neutrinos.
We, however, do not consider tensor NP couplings in our analysis. Moreover, we consider the NP coefficients to be real in our analysis. 

The three body differential decay distribution for $B_s\, \to D_s^*$ semileptonic decays can be written as
\begin{equation}
\frac{d\Gamma}{dq^2dcos\theta} = \frac{G_F^2|V_{cb}|^2|\vec P_{D_s^*}|}{2^9\pi^3 m_{B_s}^2}
{\Big(1 - \frac{m_l^2}{q^2}\Big)} L_{\mu\nu} H^{\mu \nu}\,,
\end{equation}
where, $|\vec P_{D_s^*}|$ = $\sqrt{\lambda (m_{B_s}^2, m_{D_s^*}^2,q^2)}$ is the three momentum vector of the outgoing vector meson and 
$\lambda(a,b,c)$ = $a^2+b^2+c^2-2(ab+bc+ca)$. The $q^2$ represents the lepton mass squared, $\theta_l$ represents the 
angle between $P_{D_s^*}$ and the lepton three momentum vector in the $l-\nu$ rest frame. The covariant contraction $L_{\mu\nu}
H^{\mu \nu}$ can be calculated using the helicity techniques discussed in Ref.~\cite{Korner:1989qb, Kadeer:2005aq}. 
The differential decay distribution can be expressed in terms of various helicity amplitudes as follow~\cite{Dutta:2013qaa}.
\begin{eqnarray}
\frac{d\Gamma}{dq^2dcos\theta} &=& N P_{D_s^*} \Bigg\{ 2\mathcal{A}_0^2 \sin^2\theta_l(G_A^2 + \widetilde{G}_A^2) + 
(1 + \cos^2\theta)\Big[\mathcal{A}_{||}^2(G_A^2 + \widetilde{G}_A^2) + \mathcal{A}_{\perp}^2(G_V^2 +\widetilde{G}_V^2)\Big] \nonumber \\
&&- 4\mathcal{A}_{||}\mathcal{A}_{\perp}\cos\theta_l (G_AG_V - \widetilde{G}_A\widetilde{G}_V)
+ \frac{m_l^2}{q^2}\sin^2\theta\Big[\mathcal{A}_{||}^2(G_A^2 + \widetilde{G}_A^2) + \mathcal{A}_{\perp}^2(G_V^2 + \widetilde{G}_V^2)\Big]
\nonumber \\
&&+\frac{2m_l^2}{q^2}\Big[\Big\{\mathcal{A}_0 G_A \cos\theta_l - (\mathcal{A}_t G_A +\frac{\sqrt{q^2}}{m_l}\mathcal{A}_PG_P)\Big\}^2 
+\Big\{\mathcal{A}_0 \widetilde{G}_A \cos\theta_l - (\mathcal{A}_t\widetilde{G}_A +
\frac{\sqrt{q^2}}{m_l}\mathcal{A}_P\widetilde{G}_P)\Big]\Big\}^2\Bigg\}\,,
\label{equ-1}
\end{eqnarray}
where
\begin{eqnarray}
G_V=1+V_L+V_R; \hspace{1cm} G_A=1+V_L-V_R ; \hspace{1cm} G_S=S_L-S_R \hspace{1cm} G_P=S_L-S_R\nonumber \\
\widetilde{G}_V=1+\widetilde{V}_L+\widetilde{V}_R; \hspace{1cm} \widetilde{G}_A=\widetilde{V}_L+\widetilde{V}_R;
\hspace{1cm} \widetilde{G}_S=\widetilde{S}_L+\widetilde{S}_R ; \hspace{1cm} \widetilde{G}_P=\widetilde{S}_L+\widetilde{S}_R\,
\end{eqnarray}
and
\begin{eqnarray}
&&N=\frac{G_F^2|V_{cb}|^2 q^2}{256\pi^3 m_{B_s}^2}\Big(1 - \frac{m_l^2}{q^2}\Big)^2\,, \nonumber \\
&&\mathcal{A}_0=\frac{1}{2 m_{D_s^*} \sqrt{(q^2)}}\Big[(m_{B_s}^2 - {m_{D_s^*}^2} - q^2)(m_{B_s}+m_{D_s^*})A_1(q^2)
\frac{4m_{B_s}^2{|P_{D_s^*}|}^2}{(m_{B_s}+m_{D_s^*})}A_2(q^2)\Big]\,,\nonumber \\
&&\mathcal{A}_{||}=\frac{2(m_{B_s}+m_{D_s^*})A_1(q^2)}{\sqrt{2}}\,, \qquad\qquad
\mathcal{A}_{\perp}= -\frac{4m_{B_s} V(q^2)|\vec P_{D_s^*}|}{\sqrt{2}(m_{B_s} + m_{D_s^*})}\,,\nonumber\\
&&\mathcal{A}_{t}=\frac{2m_{B_s} |P_{D_s^*}|A_0(q^2)}{\sqrt{q^2}}\,, \qquad\qquad
\mathcal{A}_{P}= - \frac{2m_{B_s} |P_{D_s^*}|A_0(q^2)}{(m_b(\mu)+m_c(\mu))}\,. 
\end{eqnarray}
Here, $V$, $A_0$, $A_1$, $A_2$ are the form form factors calculated in the relativistic quark model. We refer to Ref~\cite{Faustov:2012mt} 
for the respective form factor inputs for the $B_s\rightarrow D^{*}_{s}l\nu$ decay mode. We give prediction of several observables such as
the differential branching ratio $DBR(q^2)$, the ratio of branching ratio $R(q^2)$, the lepton polarization fraction $P^l(q^2)$, 
the forward-backward asymmetry $A_{FB}^l(q^2)$, the convexity parameter $C_F^l$, forward-backward asymmetry for the transversely polarized 
$D_s^*$ meson and the longitudinal polarization fraction $F_L^{D_s^*}$ for the $B_s\rightarrow D^{*}_{s}l\nu$ decay mode. We omit the details
as the definition of all these observables can be found elsewhere in the literature.
\section{Results and Discussions}
\label{results}
\subsection{Input parameters}
As of our theory inputs are concerned, we report in Table~\ref{inputs} the masses of various mesons, leptons and  mass of $b$ quark and $c$ quark 
evaluated at the renormalization scale $\mu = m_b$. All the mass parameters are in GeV units. The Fermi coupling constant $G_F$ is in 
GeV$^{-2}$, $|V_{cb}|$ is the corresponding CKM matrix element and the $\tau_{B_s}$ is the $B_s$ meson life time expressed in second.
We ignore the uncertainties associated with the mass parameters and the decay lifetime of $B_s$ meson. 
We consider the uncertainties associated with the CKM matrix element $|V_{cb}|$ and the form factor input parameters.
The form factor inputs, obtained in the relativistic quark model, are taken from Ref.~\cite{Faustov:2012mt}. 
The form factors inputs $V$, $A_0$, $A_1$ and $A_2$ at zero recoil $(q^2=0)$ and at the maximum recoil 
$(q^2= q^2_{max})$ and the fitted parameters $\sigma_1$ and $\sigma_2$ are reported in the Table~\ref{formfactor}.
We consider $\pm 10\%$ uncertainties in the form factor inputs. 
Similarly, for $B\,\to\, D$, $B\,\to\, D^*$ and $B_c\,\to\, {J/\psi}$ form factors, we refer to the lattice QCD results~\cite{Lattice:2015rga},
the HQET~\cite{Caprini:1997mu} and perturbative QCD (pQCD) results~\cite{Wen-Fei:2013uea}, respectively.

\begin{table}[htbp]
\centering 
\begin{tabular}{lllll}
\hline
Parameters  & Values & Parameters  & Values\\
\hline
$m_{B_s}$    & 5.36677 & $m_{D_s^*}$ & 2.1123 \\
$m_{B_c}$    & 6.272 & $m_{B^*_c}$ & 6.332 \\
$m_e$        & $0.5109989461\times10^{-3}$ & $m_\tau$ & 1.77682\\
$m_b$        & 4.18          &  $m_c$ & 0.91\\
$V_{cb}$     & 0.0409(11)   & $G_F$ & $1.1663787\times10^{-5}$ \\
$\tau_{B_s}$ & $1.516\times10^{-12}$ &  &\\
\hline
\end{tabular}
\caption{Theory input parameters~\cite{Tanabashi:2018oca} }
\label{inputs}
\end{table}

\begin{table}[htbp]
\centering 
\setlength{\tabcolsep}{6pt} % Default value: 6pt
\renewcommand{\arraystretch}{1.2}
\begin{tabular}{lllll}
\hline
& V  & $A_0$ & $A_1$  & $A_2$\\
\hline
$F(0)$ & 0.95 & 0.67 & 0.70 &0.75\\
$F(q^2_{max})$ & 1.50 & 1.06 & 0.84 & 1.04\\
$\sigma_1$ & 0.372 &0.350& 0.463 & 1.04\\
$\sigma_2$ & - 0.561 & - 0.600 & - 0.510 & - 0.070\\
\hline
\hline
\end{tabular}
\caption{$B_s \to D^{\ast}_s$ form factor inputs~\cite{Faustov:2012mt}}
\label{formfactor}
\end{table}

\subsection{SM predictions of $B_s\to D_s^* l \nu$ decay mode}

We report the SM central values and the corresponding $1\sigma$ ranges of various physical observables such as the differential 
branching ratio~($DBR$), the ratio of branching ratio~($R_{D_s^*}$), the forward backward asymmetry~($A_{FB}^l$), the convexity 
parameter~($C_F^l$), the forward backward asymmetry for the transversely polarized $D_s^*$ meson~($A_{FB}^T$) and the longitudinal 
polarization fraction of $D_s^*$ meson~($F_L^{D_s^*}$) for the $B_s \to D_s^* l \nu$ decay mode in Table~\ref{smresults}. 
The central values are obtained by considering the central values of all the input parameters and the corresponding $1\sigma$ 
ranges are obtained by performing a random scan over the theoretical inputs such as the form factors and the CKM matrix element within 
$1\sigma$ of their central values. 
The central values obtained for the branching ratio and the ratio of branching ratio is quite similar to the values reported in 
refs~\cite{Faustov:2012mt, Bhol:2014jta}. A slight difference is observed due to the different choices of input parameters. In SM, the 
branching ratio for the $B_s\rightarrow D_s^*l\nu$ decay mode is observed to be of the order of $10^{-2}$ for both $e$ and $\tau$ modes,
respectively. As expected, the longitudinal polarization fraction $P^l$ for the $e$ mode is $-\,1.00$. 

\begin{table}[htbp]
\centering
\setlength{\tabcolsep}{6pt} % Default value: 6pt
\renewcommand{\arraystretch}{1.5} % Default value: 1
\begin{tabular}{|c|c|c|c|c|c|c|}
\hline
\hline
Observable &\multicolumn{2}{c|}{$e$ mode } &\multicolumn{2}{c|}{$\tau$  mode} \\
\cline{2-5}
& Central value & $1\sigma$ range & Central value & $1\sigma$ range\\
\hline
\hline
{$DBR\times10^{-2} $} & 5.92 &  (5.37, 6.49) & 1.42 & (1.29, 1.56)  \\
\hline
$A_{FB}^{l}$        & -0.256 &(-0.269, -0.244) &-0.087 & (-0.097, -0.078)\\
\hline
$P^{l}$             & -1.000 & -1.000  & -0.523 & (-0.532, -0.514)\\
\hline
$C_F^l$             & -0.362 & (-0.385, -0.339) & -0.042 & (-0.048, -0.036)\\
\hline
$A_{FB}^{T}$        & -0.507 & (-0.521, -0.490) & -0.356 & (-0.369, -0.343) \\
\hline
$F_L^{D^*}$         & 0.494  & (0.484, 0.504) & 0.431 & (0.425, 0.437) \\
\hline
\hline
$R_{D_s^*}$ & \multicolumn{2}{c|}{0.241}  & \multicolumn{2}{c|}{(0.238, 0.244)}\\
\hline
\hline
\end{tabular}
\caption{Central values and the corresponding $1\sigma$ ranges of various observables in the SM for the
$B_s\to D_s^*l\nu$ decay mode.}
\label{smresults}
\end{table}

\begin{figure}[htbp]
\centering
\includegraphics[width=4cm,height=3cm]{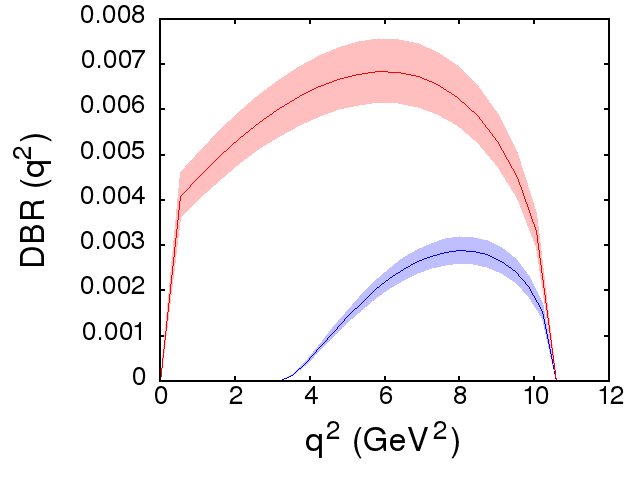}  
\includegraphics[width=4cm,height=3cm]{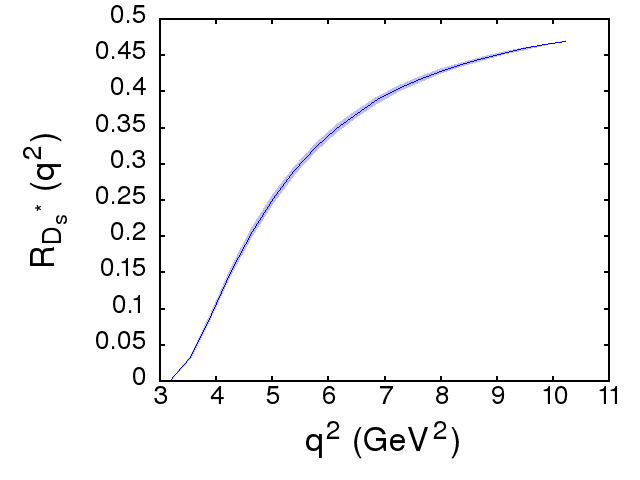}
\includegraphics[width=4cm,height=3cm]{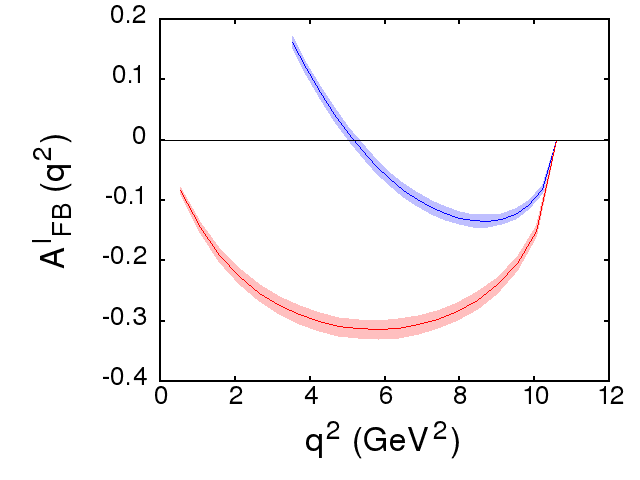}
\includegraphics[width=4cm,height=3cm]{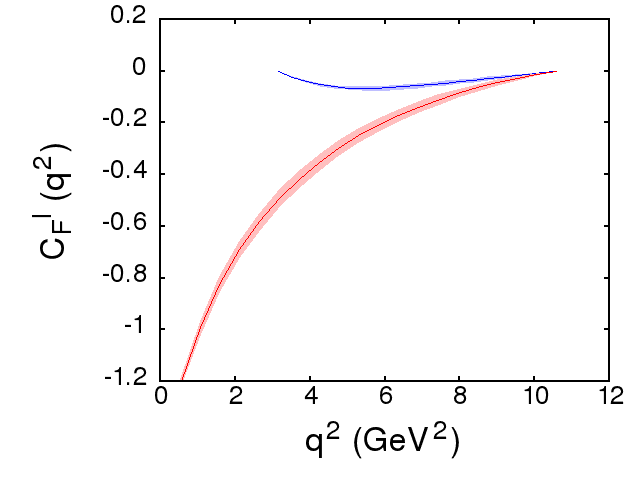}
\includegraphics[width=4cm,height=3cm]{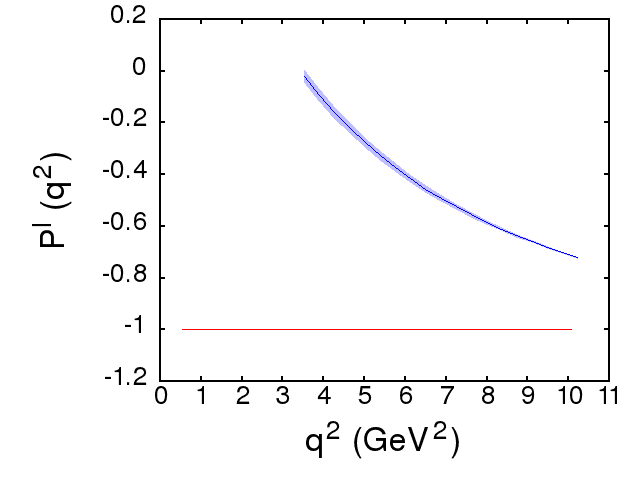}
\includegraphics[width=4cm,height=3cm]{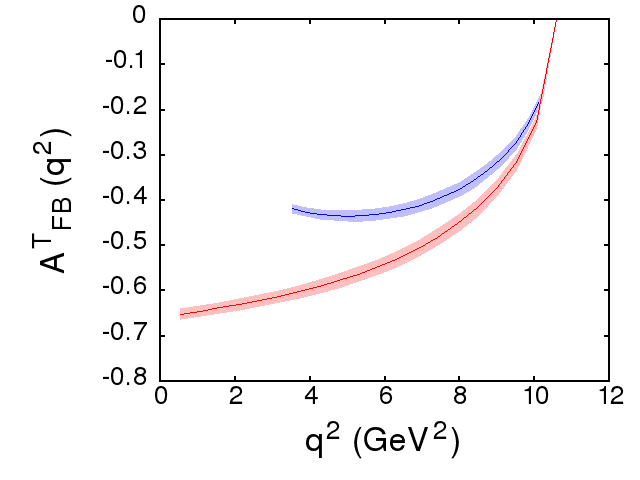}
\includegraphics[width=4cm,height=3cm]{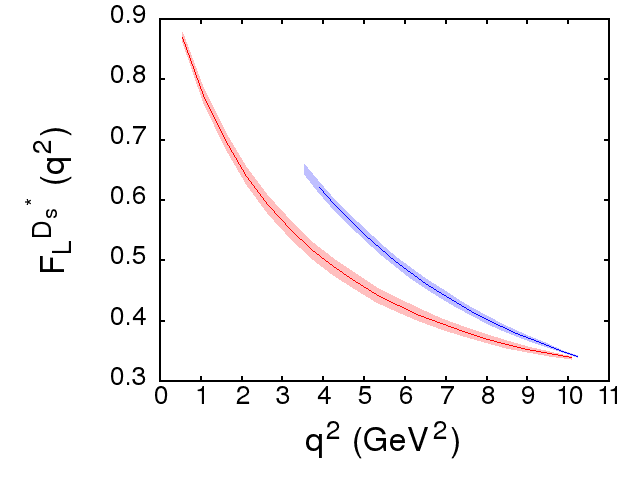}
\caption{$q^2$ dependence of $B_s \to D_s^*l\nu$ decay observables in the SM for the $e$~(red) and the $\tau$~(blue) mode. }
\label{smplots}
\end{figure}

In Fig~\ref{smplots}, we show the $q^2$ dependent plots for each observables for the $B_s \to D_s^{\ast}\,l\,\nu$ decays. The solid line
corresponds to the central value of each input parameters
and the band corresponds to the uncertainties associated with $|V_{cb}|$ and $B_s\to D_s^*l\nu$ form factor inputs. The blue solid 
line~(band) represents the $\tau$ mode and the red solid line~(band) represents the $e$ mode. Our observations are as follows:
\begin{itemize}

\item Uncertainty associated with all the observables is much less compared to the differential branching ratio. This is expected as the 
uncertainties associated with the CKM matrix element and the form factor inputs get cancelled to some extent in these ratios.

\item The  differential decay distribution is zero at the zero recoil and the maximum recoil points.
The peak of the differential decay 
distribution is observed at $q^2 \approx 5.9\,{\rm GeV^2}$ for the $e$ 
mode and at $q^2\approx 8.1\,{\rm GeV^2}$ for the $\tau$ mode. 
The ratio of branching ratio is maximum, i.e, $R\,(q^2) \approx  0.46$ at $q^2 = q^2_{max}$.

\item The forward backward asymmetry $A^e_{FB}$ is negative for the whole $q^2$ region, whereas, 
$A^{\tau}_{FB}\,(q^2)$ has a peak at low $q^2$ and 
it gradually decreases as $q^2$ increases.
We also observe a zero crossing in $A_{FB}^{\tau}\,(q^2)$ at $q^2\approx 5.2\pm 0.2\,{\rm GeV^2}$. 

\item The convexity parameter, $C_F^e\,(q^2)$ is found to have negative values for the whole $q^2$ range and it increases as $q^2$ increases. At
$q^2 = q^2_{\rm max}$, the convexity parameter becomes equal to zero  for both $e$ and $\tau$ mode, respectively. 

\item The polarization fraction, $P^e(q^2)$ is constant over entire $q^2$ region whereas $P^{\tau}(q^2)$ decreases as $q^2$ increases.

\item The forward backward asymmertry $A_{FB}^T\,(q^2)$ for the transversely polarized 
$D_s^*$ meson is negative in the whole $q^2$ region for both $e$ mode and $\tau$ mode. Also, it gradually increases as the $q^2$ increases and 
becomes zero at maximum value of $q^2$. Similarly, the polarization fraction of $D_s^*$ meson $F_L^{D_s^*}\,(q^2)$ is observed to have maximum value 
at low $q^2$ for both $\tau$ mode and $e$ mode and it gradually decreases as $q^2$ increases.

\end{itemize}

\subsection{$\chi^2$ analysis}
% \subsubsection{NP predictions on $B\to(D,D^*)l\nu$ and $B_c\to J/\psi l\nu$ decay mode}
\label{sec:Alice} 
Our main objective is to investigate the anomalies present in $b \to c\tau\nu$ decays within a model independent framework and to find the 
minimal 
number of NP couplings that best fit the data. First, to obtain the amount of discrepancy of SM with the experimental data, we perform a 
naive $\chi^2$ analysis defined as
\begin{equation}
\chi^2=\sum_{i} \frac{{(\mathcal{O}_i^{th} - \mathcal{O}_i^{exp})}^2}{({\Delta \mathcal{O}_i^{exp}})^2}\,,
\end{equation}
where $\mathcal{O}_i^{th}$ and $\mathcal{O}_i^{exp}$ refer to the theoretical and the experimental values of $R_D$, $R_{D^*}$, $R_{J/\psi}$, 
$P_{\tau}^{D^*}$ and $F_L^{D^*}$. $\Delta \mathcal{O}_i^{exp}$ represents the corresponding experimental uncertainties associated with $R_D$,
$R_{D^*}$, $R_{J/\psi}$, $P_{\tau}^{D^*}$ and $F_L^{D^*}$. The total $\chi^2$ is evaluated by including the five measurements as 
mentioned above. The $\chi^2_{min}$ for SM is evaluated by performing a random scan over the form factor input parameters 
and the CKM matrix element within $1\sigma$ of their central values. Similarly, the best fit values of all the NP couplings such as 
$V_L$, $V_R$, $S_L$, $S_R$, $\widetilde{V}_L$, $\widetilde{V}_R$, $\widetilde{S}_L$ and $\widetilde{S}_R$ are also obtained. We report in 
Table~\ref{nprd} and ~\ref{npafb} the best estimates of all the observables pertaining to $B \to (D,\,D^{\ast})\tau\nu$ and 
$B_c \to J/\Psi\tau\nu$ decays by considering NP couplings one at a time and we call it as the 1D scenario.

 \begin{table}[htbp]
 \centering
  \setlength{\tabcolsep}{6pt} % Default value: 6pt
\renewcommand{\arraystretch}{1.5} % Default value: 1
\begin{tabular}{l l l l l l l l l l l}
\hline
Coefficient & Best fit value & $R_D$ & $R_{D^*}$ & $R_{J/\psi}$ & $\mathcal{B}(B_c\rightarrow\tau\nu)\%$ & $P_{\tau}^{D^*}$&  $P_{\tau}^{J/\psi}$& $F_{L}^{D^*}$  & $F_{L}^{J/\psi}$ & ${\chi}^{2}_{min}$ \\
\hline

SM  & & 0.332 & 0.255 & 0.288 & 1.959 & -0.501 & -0.472 & 0.455 & 0.420 & 14.3\\

$V_L$ & 0.087 & 0.340 & 0.297 & 0.344 & 2.164 & -0.493 & -0.454 & 0.462 & 0.427 & 4.8 \\
 
$V_{R}$ & -0.063 & 0.290 & 0.282 & 0.326 & 2.043 & -0.492 & -0.465 & 0.467 & 0.424 & 8.6\\

$S_{L}$ & 0.001 &  0.326 & 0.255 & 0.289 & 2.037 & -0.500 & -0.468 & 0.455 & 0.421 & 14.6\\
        
$S_R$  & 0.211 & 0.360 & 0.262 & 0.299 & 4.072 & -0.456 & -0.420 & 0.472 & 0.441 & 11.4  \\
 
${\widetilde{V}}_L$ & 0.418 & 0.337 & 0.295 & 0.341 & 2.318 & -0.347 & -0.323 & 0.462 & 0.425 & 4.8  \\

${\widetilde{V}}_R$ & 0.418 & 0.337 & 0.295 & 0.341 & 2.318 & -0.347 & -0.323 & 0.462 & 0.425 & 4.8 \\
 
${\widetilde{S}}_L$ & 0.576 & 0.360 & 0.259 & 0.294 & 15.023 & -0.508 & -0.476 & 0.464 & 0.432 & 12.6\\
 
${\widetilde{S}}_R$ &  0.576 & 0.360 & 0.259 & 0.294 & 15.023 & -0.508 & -0.476 & 0.464 & 0.432 &12.6 \\
\hline
\hline
\end{tabular}
\caption{Best fit values of $R_D$, $R_D^*$, $R_{J/\psi}$,$\mathcal{B}(B_c\rightarrow \tau\nu)\%$ , $P_\tau^{D^*}$, $F_L^{D^*}$, $P_\tau^{J/\psi}$ and  
$F_L^{J/\psi}$ within SM and in the presence of various NP couplings in $1D$ scenario.}
\label{nprd}
\end{table}

\begin{table}[htbp]
\centering
\setlength{\tabcolsep}{4pt} % Default value: 6pt
\renewcommand{\arraystretch}{1.5} % Default value: 1
\begin{tabular}{l l l l l l l l l l l l l}
\hline
Coefficient & Best fit value & $P_{\tau}^{D}$ & $A_{FB}^{\tau}(D)$&  $A_{FB}^{\tau}(D^*)$ & $A_{FB}^{\tau}(J/\psi)$ & $A_{FB}^{T( D^*)}$ & $A_{FB}^{T (J/\psi)}$ &$C_F^{\tau(D)}$& $C_F^{\tau(D^*)}$ & $C_F^{\tau(J/\psi)}$\\
%   & ${\chi}^{2}_{min}$ \\
 \hline
 
  SM &       &    0.352  &  0.358   &    -0.063    &    0.011   &    -0.352       & -0.199   &    -0.260      & -0.056   &    -0.012   \\
 
 $V_L$      &    0.087  &    0.311    &    0.361   &    -0.057      &  0.018    &   -0.352   &    -0.198    &   -0.275  &     -0.059    &   -0.011      \\
  
 $V_{R}$    &   -0.063   &     0.347  &      0.358   &    -0.037  &      0.028  &     -0.321   &    -0.175   &    -0.262  &     -0.063   &    -0.013   \\
 
 $S_{L}$    &  0.001      &  0.355       & 0.357  &     -0.067     &   0.015     &  -0.361  &     -0.195 &      -0.259 &      -0.055  &     -0.012  \\
  
 $S_R$     &  0.211   &     0.478  &      0.335    &   -0.083 &      -0.002 &      -0.358    &   -0.191   &    -0.208  &     -0.054  &     -0.012  \\
  
 ${\widetilde{V}}_L$   &    0.418  &      0.226  &      0.360   &    -0.001 &       0.053      & -0.249    &   -0.135  &     -0.271  &     -0.059  &     -0.012        \\
 
 ${\widetilde{V}}_R$      & 0.418    &    0.226      &  0.360  &     -0.057    &    0.020      & -0.354   &    -0.192     &  -0.271   &    -0.059  &     -0.012  \\
  
 ${\widetilde{S}}_L$      &   0.576  &     -0.039   &     0.268   &    -0.062     &   0.013 &      -0.352   &    -0.199    &   -0.209  &     -0.055    &   -0.011   \\
         
 ${\widetilde{S}}_R$       &     0.576   &    -0.039   &     0.268   &    -0.062 &       0.013   &    -0.352   &    -0.199    &   -0.209   &    -0.055    &   -0.011 \\
 \hline     
 \hline
  \end{tabular}
\caption{Best fit values of $P_{\tau}^D$, $A_{FB}^{\tau}(D)$, $A_{FB}^{\tau}(D^*),A_{FB}^{\tau}(J/\psi),A_{FB}^{T}(D^*),A_{FB}^{T}(J/\psi), F_L^{D^*},
F_L^{J/\psi}, C_F^{\tau(D)}, C_F^{\tau(D^*)},C_F^{\tau(J/\psi)}$  within SM and  in the presence of various NP couplings in $1D$ scenario}
\label{npafb}
\end{table}
We obtain the $\chi^2_{min}$ for SM to be 14.3. It should be noted that with $S_L$ NP coupling the fit worsens as the $\chi^2_{min}$ 
obtained in this scenario is more than the value obtained in the SM. The best estimates of  
$\mathcal{B}(B_c\rightarrow \tau\nu)$ in SM is found to be $1.959\%$ and it is in good agreement with other
predictions~\cite{Akeroyd:2017mhr}. It is worth emphasizing that in the presence of the new scalar NP couplings such as
${\widetilde{S}}_L$ and ${\widetilde{S}}_R$, we obtain $\mathcal{B}(B_c\rightarrow \tau\nu)$ to be more than $10\%$,
the upper bound of $\mathcal{B}(B_c\rightarrow \tau\nu)$ estimated in the SM~\cite{Akeroyd:2017mhr}. Again, the minimum $\chi^2$ obtained 
is rather large in case of $V_R$ and $S_R$ NP couplings.
Hence a simultaneous explanation of the anomalies present in $R_D$, $R_{D^*}$, $R_{J/\psi}$, $P_\tau^{D^*}$ and
$F_L^{D^*}$ can be found with $V_L$, $\widetilde{V}_L$ and $\widetilde{V}_R$ NP couplings.

We now consider the NP contributions by considering two different NP couplings at a time. We report four such 2D NP scenarios, namely 
($V_L$, $V_R$), ($\widetilde{V}_L$, $\widetilde{ V}_R$), ($S_L$, $S_R$) and ($\widetilde{S}_L$, $\widetilde{S}_R$).
In Table.\ref{2drd} and ~\ref{2drd2} we give the best estimates  of these NP couplings and the corresponding best estimates of 
all the observables pertaining to $B \to (D,\,D^{\ast})\tau\nu$ and $B_c \to J/\Psi\tau\nu$ decays.
In the 2D scenarios, the $\chi^2_{min}$ reduces significantly from that of $1D$ scenarios. We observe that with ($S_L$, $S_R$) NP
couplings we obtain the best fit to the data with $\chi^2_{\rm min} = 2.6$. However, it produces $\mathcal{B}(B_c\to \tau\nu)$ which is
slightly more than the $10\%$ upper bound obtained in the SM. The scenarios with $({S}_L, {S}_R)$ and 
($\widetilde{S}_L$, $\widetilde{S}_R$) are strongly disfavored as the best estimates of $\mathcal{B}(B_c\to \tau\nu)$ obtained in these 
scenarios are more than the total decay width of $B_c$ meson. Scenarios with ($V_L$, $V_R$) and($\widetilde{V}_L$, $\widetilde{V}_R)$ 
NP couplings are consistent with the $\mathcal{B}(B_c\to \tau\nu)$ constraint.

\begin{table}[htbp]
\label{2d1}
\centering
\setlength{\tabcolsep}{6pt} % Default value: 6pt
\renewcommand{\arraystretch}{1.5} % Default value: 1
\begin{tabular}{l l l l l l l l l l l}
\hline
Coefficient & Best fit value & $R_D$ & $R_{D^*}$ & $R_{J/\psi}$ & $\mathcal{B}(B_c\rightarrow\tau\nu)\%$ & $P_{\tau}^{D^*}$ & $P_{\tau}^{J/\psi}$ &$F_{L}^{D^*}$ &$F_{L}^{J/\psi}$ & ${\chi}^{2}_{min}$  \\
\hline
$(V_L, V_R$)  & (0.087, -0.004) & 0.343 & 0.299 & 0.347 & 2.276 & -0.493 & -0.457 & 0.461 & 0.426 & 4.8\\

($S_L, S_R$)  & (-0.467, 0.573) & 0.342 & 0.300 & 0.361 & 10.835 & -0.256 & -0.174 & 0.546& 0.537 & 2.6\\

$(\widetilde{V}_L,\widetilde{V}_R)$ & (-0.350, 0.091) & 0.338 & 0.299 & 0.345 & 2.344 & -0.339 & -0.316 & 0.464 & 0.422 & 4.7\\

$(\widetilde{S}_L,\widetilde{S}_R)$ & (-0.966, 0.953) & 0.328 & 0.297 & 0.352 & 163.236 & -0.573 & -0.553 & 0.541 & 0.525 & 2.8\\
\hline
\hline
\end{tabular}
\caption{Best fit values of $R_D$, $R_D^*$, $R_{J/\psi}$, $\mathcal{B}(B_c\rightarrow \tau\nu)\%$, $P_\tau^{D^*}$, $F_L^{D^*}$, $P_\tau^{J/\psi}$ and  
$F_L^{J/\psi}$ in presence of NP in $2D$ scenario.}
\label{2drd}
\end{table}

\begin{table}[htbp]
\label{2d2}
\centering
\setlength{\tabcolsep}{4pt} % Default value: 6pt
\renewcommand{\arraystretch}{1.5} % Default value: 1
\begin{tabular}{l l l l l l l l l l l l l l}
\hline
Coefficient & Best fit value & $P_{\tau}^D$ & $A_{FB}^{\tau}(D)$&  $A_{FB}^{\tau}(D^*)$ & $A_{FB}^{\tau}(J/\psi)$ & $A_{FB}^{T( D^*)}$ & $A_{FB}^{T (J/\psi)}$ &$C_F^{\tau(D)}$& $C_F^{\tau(D^*)}$ & $C_F^{\tau(J/\psi)}$\\
 
 \hline

 $(V_L,V_R)$  & ( 0.087, -0.004) & 0.317 & 0.360 & -0.061 & 0.019 & -0.360 & -0.195 & -0.273 & -0.058 &  -0.012\\ 
 
($S_L,S_R$)  &  (-0.467, 0.573) &  0.418  & 0.346  & -0.137  & -0.066 &  -0.356 & -0.197  & -0.233 & -0.048 & -0.009\\ 

($\widetilde{V}_L, \widetilde{V}_R $) & (-0.350, 0.091) & 0.288 & 0.359 & -0.010 & 0.042  &  -0.268 & -0.149 & -0.269 & -0.061 & -0.012 \\

($\widetilde{S}_L, \widetilde{S}_R $) & (-0.966, 0.953) & 0.341 & 0.358 & -0.054 & 0.012 & -0.359 & -0.201 & -0.265 & -0.048 & -0.009 \\

\hline     
\hline
\end{tabular}
\caption{Best fit values for of $P_{\tau}^D$, $A_{FB}^{\tau}(D)$, $A_{FB}^{\tau}(D^*),A_{FB}^{\tau}(J/\psi),A_{FB}^{T}(D^*),A_{FB}^{T}(J/\psi), F_L^{D^*},
F_L^{J/\psi}, C_F^{\tau(D)}, C_F^{\tau(D^*)},C_F^{\tau(J/\psi)}$ in the presence of NP in $2D$ scenario.}
\label{2drd2}
\end{table}

\subsection{$B_s\to D_s^* \tau\nu$ decay observables in $1D$ and $2D$ scenarios} 

\subsubsection{1D scenario}
Our objective here is to see the effect of NP on various observables pertaining to $B_s\to D_s^* \tau\nu$ decays. In Table.~\ref{npbsds}, we 
report the best estimates of all the observables obtained using the best fit values of various NP couplings of Table.~\ref{nprd}.
We see a significant deviation in $R_{D_s^{\ast}}$ from the SM prediction with vector NP couplings. However,
the deviation observed is quite negligible with scalar NP couplings. The deviation observed in $P_{\tau}^{D_s^{\ast}}$ is more pronounced with
$\widetilde{V}_L$ and $\widetilde{V}_R$ NP couplings. Similarly, maximum deviation from the SM prediction is observed with $\widetilde{V}_L$ 
NP coupling in case of $A_{FB}^{\tau}$ and $A_{FB}^T$ observables.   
\begin{table}[htbp]
\centering
 \setlength{\tabcolsep}{6pt} % Default value: 6pt
\renewcommand{\arraystretch}{1.5} % Default value: 1
\begin{tabular}{l l l l l l l l l l l l}
\hline
 Coefficient & Best fit value & $R_{D_s^*}$ & $DBR\%$  & $P_{\tau}^{D_s^*}$ & $F_{L}^{D_s^*}$ & $A_{FB}^{\tau}$ & $A_{FB}^T$ & $C_F^{\tau}$ & ${\chi}^{2}_{min}$  \\
 \hline
 
SM         & & 0.240 & 1.374 & -0.520 & 0.433 & -0.084 & -0.355 & -0.043 & 14.3\\

$V_L$      & 0.087 & 0.284 & 1.618 & -0.521 & 0.431 & -0.089 & -0.360 & -0.041 & 4.8\\
  
$V_{R}$    & -0.063 & 0.269 & 1.483 & -0.521 & 0.436 & -0.066 & -0.326 & -0.046 & 8.6 \\
 
$S_{L}$    & 0.001 & 0.241  & 1.403 & -0.519 & 0.428 & -0.093 & -0.366 & -0.038 & 14.6 \\
  
$S_R$      & 0.211 & 0.247 & 1.497 & -0.484 & 0.443 & -0.104 & -0.360 & -0.039 & 11.4\\
  
${\widetilde{V}}_L$ & 0.418 & 0.282 & 1.660 & -0.368 &  0.435 & -0.022 & -0.244 & -0.046 & 4.8 \\
 
${\widetilde{V}}_R$ & 0.418 & 0.282 & 1.660 & -0.368 & 0.435 & -0.080 & -0.347 & -0.046 & 4.8\\
  
${\widetilde{S}}_L$ & 0.576 & 0.242 & 1.374 & -0.529 & 0.441 & -0.084 & -0.354 & -0.044 & 12.6 \\
  
${\widetilde{S}}_R$ & 0.576 & 0.242 & 1.374 & -0.529 & 0.441 & -0.084 & -0.354 & -0.044 & 12.6 \\
\hline
\hline
\end{tabular}
\caption{Best estimates of $R_{D_s^*}$, $DBR\%$, $P_{\tau}^{D_s^*}$, $F_{L}^{D_s^*}$, $A_{FB}^{\tau}$, $A_{FB}^T$ and $C_F^{\tau}$ 
for $B_s\rightarrow D_s^{*}\tau\nu$ decay mode within the SM and within various $1D$ NP scenarios.}
\label{npbsds}
\end{table}

\begin{figure}[htbp]
\centering
\includegraphics[width=8.9cm,height=5.5cm]{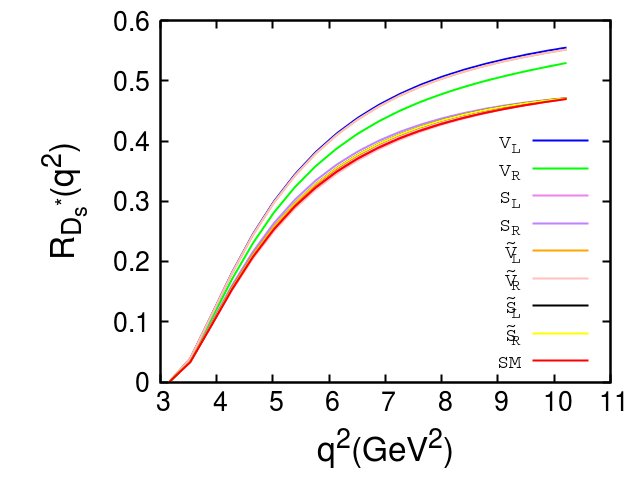}
\includegraphics[width=8.9cm,height=5.5cm]{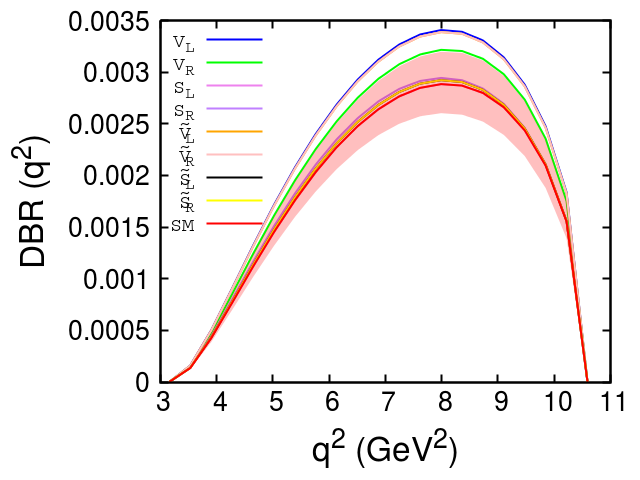}
\includegraphics[width=8.9cm,height=5.5cm]{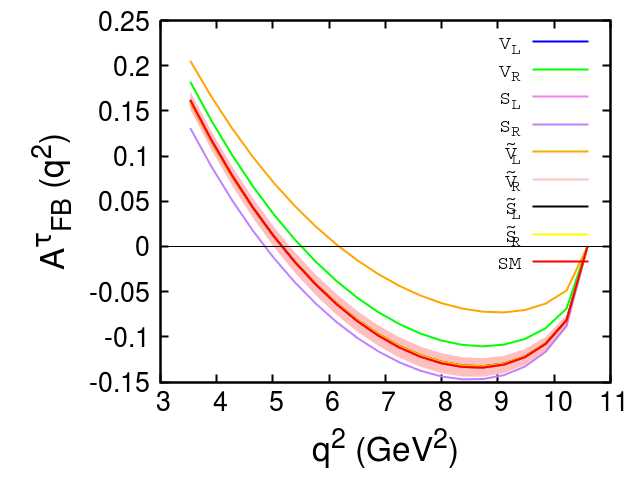}
\includegraphics[width=8.9cm,height=5.5cm]{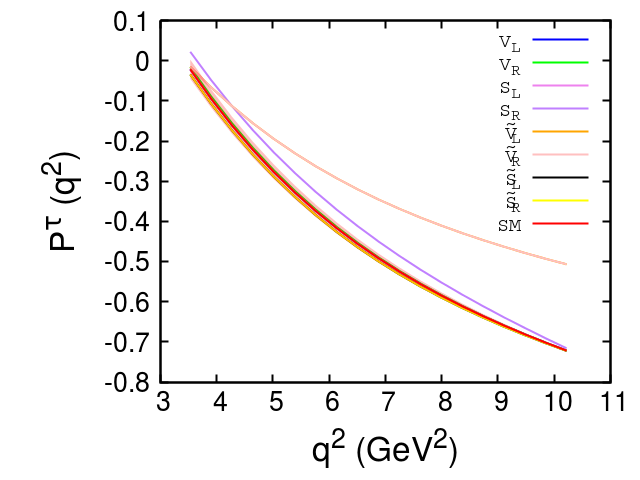}
\includegraphics[width=8.9cm,height=5.5cm]{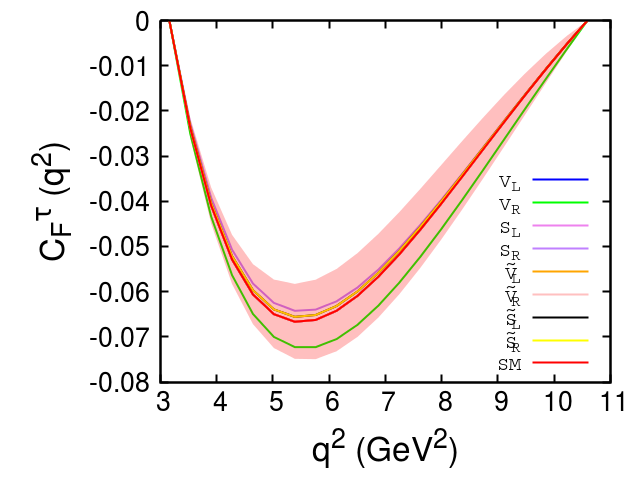}
\includegraphics[width=8.9cm,height=5.5cm]{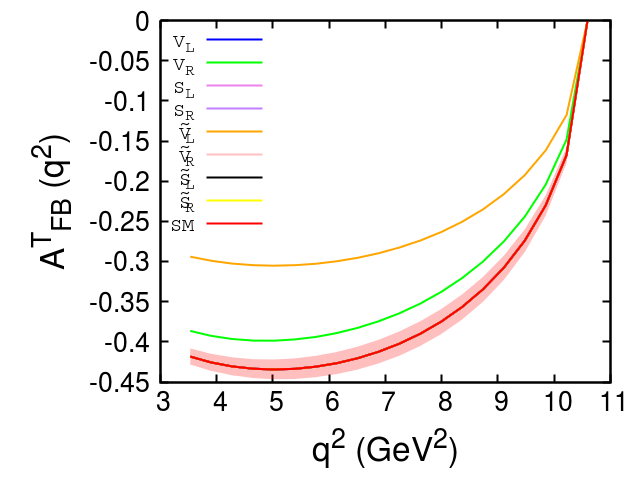}
\includegraphics[width=8.9cm,height=5.5cm]{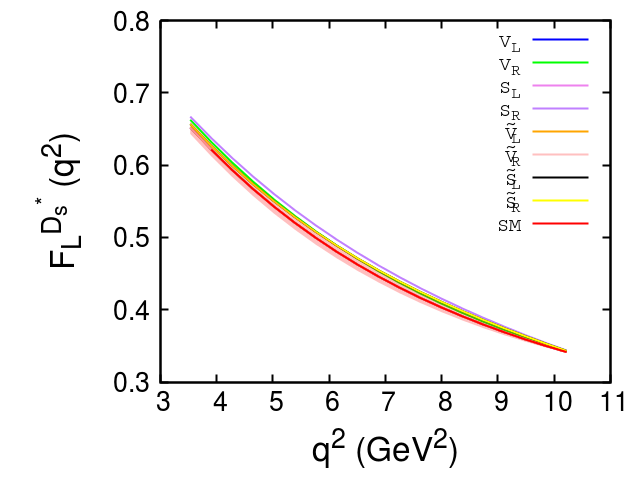}
\caption{$q^2$ dependency of $R_{D_s^*}(q^2)$, $DBR(q^2)$, $A_{FB}^{\tau}(q^2)$,  $P_{\tau}^{D_s^*}(q^2)$, 
$A_{FB}^T(q^2)$, $C_F^{\tau}(q^2)$ and $F_{L}^{D_s^*}(q^2)$ for $B_s\rightarrow D_s^* {\tau} \nu$ decay mode in the SM and in several NP 
cases. The SM central curve and the error band are shown with red colour. The corresponding best fits for $V_L$ (blue), $V_R$ (green), 
$S_L$ (violet), $S_R$ (purple), 
$\widetilde{V}_L$ (orange), $\widetilde{V}_R$ (pink), $\widetilde{S}_L$ (black), $\widetilde{S}_R$ (yellow)
are also shown. }
\label{npplots}
\end{figure}

In Fig.~\ref{npplots}, we show the $q^2$ dependency of all the observables in the SM and in various NP cases.
The SM central curve is represented by the red solid line. 
The corresponding best fits with NP couplings $V_L$~(blue), $V_R$~(green), $S_L$~(violet), $S_R$~(purple), $\widetilde{V}_L$~(orange), 
$\widetilde{V}_R$~(pink), $\widetilde{S}_L$~(black) and $\widetilde{S}_R$~(yellow) are also plotted along with the SM. The observations
pertaining to the $B_s\rightarrow D_s^*{\tau}\nu$ decay mode are as follows.
\begin{itemize}  
\item The deviation of $R_{D_s^*}(q^2)$ and $DBR(q^2)$ from their SM prediction is more pronounced and they are clearly distinguishable from
the SM prediction in case of $V_L$ and $V_R$ NP couplings. 
However, all the other NP couplings are not distinguishable from the SM prediction as $R_{D_s^*}(q^2)$ and $DBR(q^2)$ obtained with
these NP couplings lie within the SM error band. 

\item The SM zero crossing in $A^{\tau}_{FB}(q^2)$ is observed at $q^2 \approx 5.2\pm 0.2\,{\rm GeV^2}$. It, however, shifted to a 
lower value
of $q^2 \approx 4.9\,{\rm GeV^2}$ with $S_R$ NP coupling. Similarly, the zero crossing shifted to higher values of 
$q^2 \approx 5.5 \,{\rm GeV^2}$ and $6.2\,{\rm GeV^2}$ with $V_R$ and $\widetilde{V}_L$ NP couplings, respectively. They are clearly
distinguishable from the SM prediction at around $1.5\sigma$ and $5\sigma$ significance. For all other
NP couplings the zero crossing point is indistinguishable from the SM prediction.

\item The deviation of $P^{\tau}(q^2)$ from the SM prediction is more pronounced in case of $\widetilde{V}_L$ and $\widetilde{V}_R$ NP
couplings and they are clearly distinguishable from the SM prediction. Similarly, the deviation observed in case of $S_R$ NP coupling is
also distinguishable from the SM prediction. The NP effect coming from $V_L$, $V_R$, $S_L$, $\widetilde{S}_L$, 
and $\widetilde{S}_R$ NP couplings, however, is quite negligible. In case of $C_F^{\tau} (q^2)$, no significant deviation from the SM
prediction is observed.

\item In case of $A_{FB}^T (q^2)$, we see significant deviation from the SM prediction once 
$V_R$ and $\widetilde{V}_L$ NP couplings are switched on. Although, with $\widetilde{V}_L$ NP coupling, we see maximum deviation, both $V_R$ 
and $\widetilde{V}_L$ NP couplings are 
clearly distinguishable from the SM prediction. In case of $F_L^{D_s^{\ast}}$, the deviation observed from the SM prediction with
most of the NP couplings lies within the SM error band. It, however, lies slightly above the SM $1\sigma$ error band
in case of $S_R$ NP coupling.
\end{itemize} 
 
\subsubsection{2D scenario}
In the Table.\ref{2dbsdst}, we report the best estimates of all the observables for $B_s\to D_s^*\tau\nu$ decay mode by
considering two NP couplings at a time. Here we consider four different 2D scenarios such as
($V_L, V_R$), ($S_L, S_R$), ($\widetilde{V}_L, \widetilde{V}_R $) and ($\widetilde{S}_L, \widetilde{S}_R $). We see significant
deviation of $R_{D_s^{\ast}}$ from the SM prediction in each scenarios. Similar conclusion can be made for the $B_s \to D_s^{\ast}\tau\nu$
branching ratio as well. In case of $P_{\tau}^{D_s^{\ast}}$ and $A_{FB}^{\tau}$, the deviation observed is more pronounced with 
($S_L, S_R$) and ($\widetilde{V}_L, \widetilde{V}_R $) NP scenarios. Similarly, we see significant deviation in $F_L^{D_s^{\ast}}$ from
the SM prediction in case of ($S_L, S_R$) and ($\widetilde{S}_L, \widetilde{S}_R $) NP scenarios. It should be noted that only 
with ($\widetilde{V}_L, \widetilde{V}_R $) NP couplings, $A_{FB}^T$ changes appreciably from the SM prediction.
 
\begin{table}[htbp]
\label{2dds}
\centering
\setlength{\tabcolsep}{6pt} % Default value: 6pt
\renewcommand{\arraystretch}{1.5} % Default value: 1
\begin{tabular}{l l l l l l l l l l l l}
\hline
Coefficient & Best fit value &  \textbf{$R_{D_s^*}$} & $DBR\%$  & \textbf{$P_{\tau}^{D_s^*}$}& \textbf{$F_{L}^{D_s^*}$}& 
\textbf{$A_{FB}^{\tau}$}&\textbf{$A_{FB}^T$}&\textbf{$C_F^{\tau}$} &  \textbf{${\chi}^{2}_{min}$}  \\
\hline
($V_L$, $V_R$)   &  (0.087, -0.004) & 0.288 & 1.725 & -0.517 &  0.432 & -0.086 & -0.358 & -0.041 & 4.8\\
($S_L$, $S_R$)   & (-0.467, 0.573)  & 0.280 & 1.592 & -0.326 &  0.504 & -0.151 & -0.352 & -0.036 & 2.6\\
($\widetilde{V}_L$, $\widetilde{V}_R $) & (-0.350, 0.091) & 0.284 & 1.624 & -0.360 &  0.437 & -0.032 & -0.264 & -0.048 & 4.7\\
($\widetilde{S}_L$, $\widetilde{S}_R $) &(-0.966, 0.953) & 0.275 & 1.692 & -0.582 &   0.499 & -0.080 & -0.362 & -0.036 & 2.8\\ 
\hline
\hline
\end{tabular}
\caption{Best fit values of $B_s\rightarrow D_s^{*}{\tau}\nu$ decay observables $R_{D_s}$, $DBR\%$, $P_{\tau}^{D_s^*}$, $F_{L}^{D_s^*}$, 
$A_{FB}^{\tau}$, $A_{FB}^T$, $C_F^{\tau}$ within several $2D$ NP scenarios.}
\label{2dbsdst}
\end{table}

\begin{figure}[htbp]
\centering
\includegraphics[width=8.9cm,height=5.5cm]{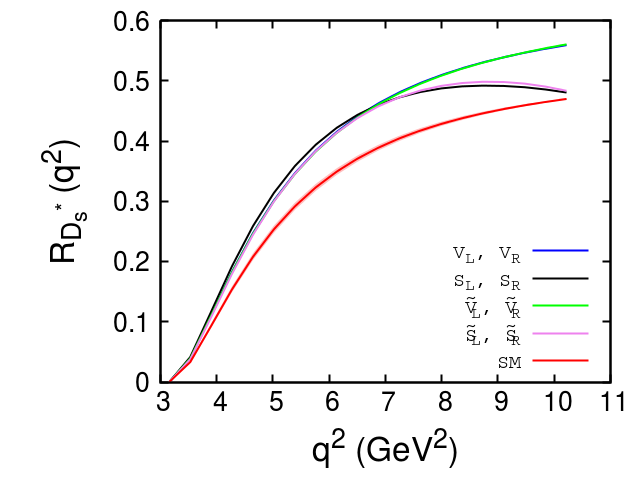}
\includegraphics[width=8.9cm,height=5.5cm]{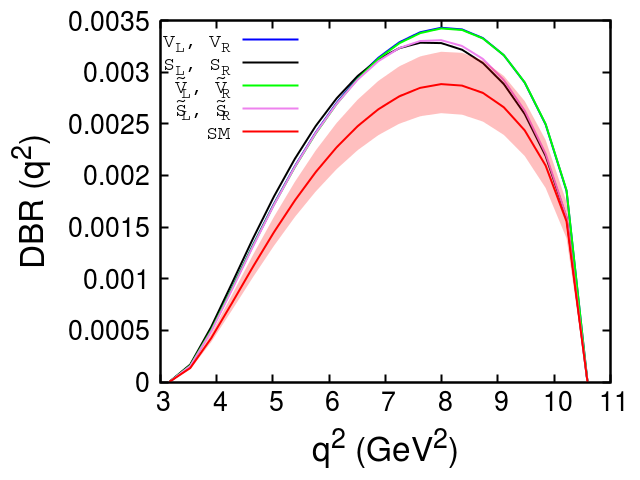}
\includegraphics[width=8.9cm,height=5.5cm]{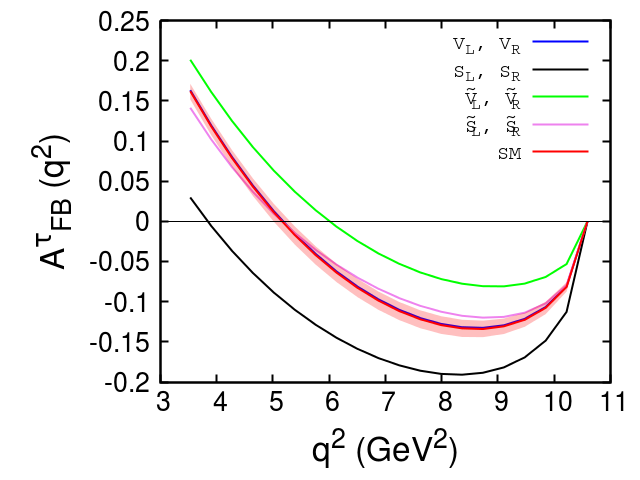}
\includegraphics[width=8.9cm,height=5.5cm]{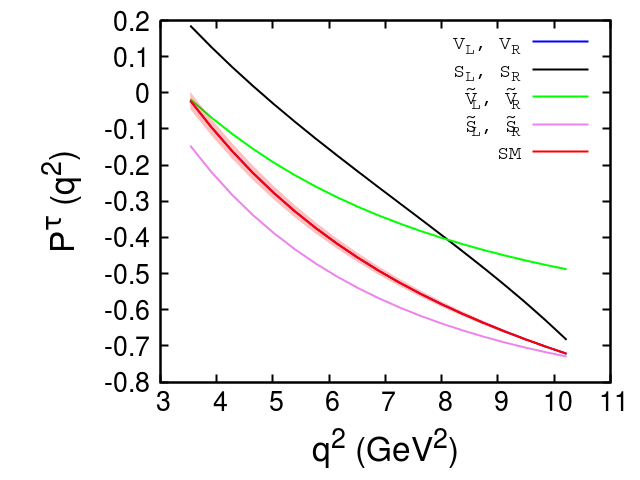}
\includegraphics[width=8.9cm,height=5.5cm]{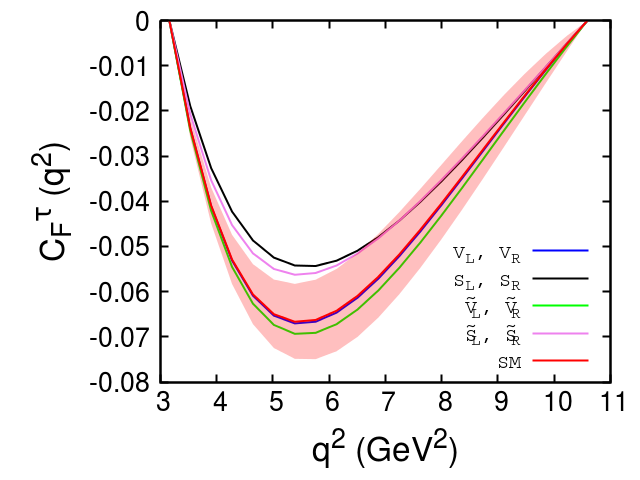}
\includegraphics[width=8.9cm,height=5.5cm]{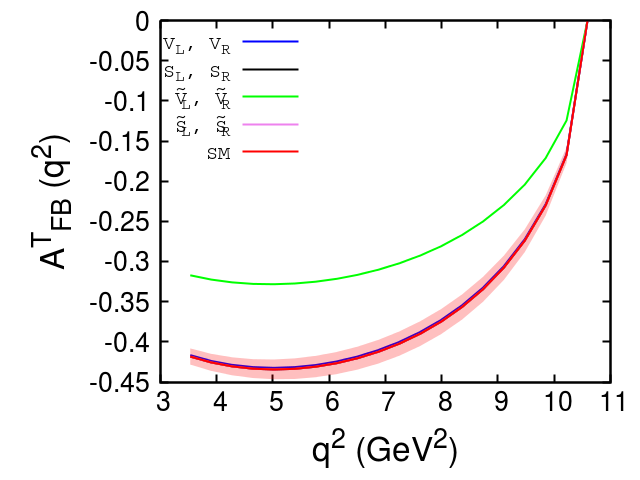}
\includegraphics[width=8.9cm,height=5.5cm]{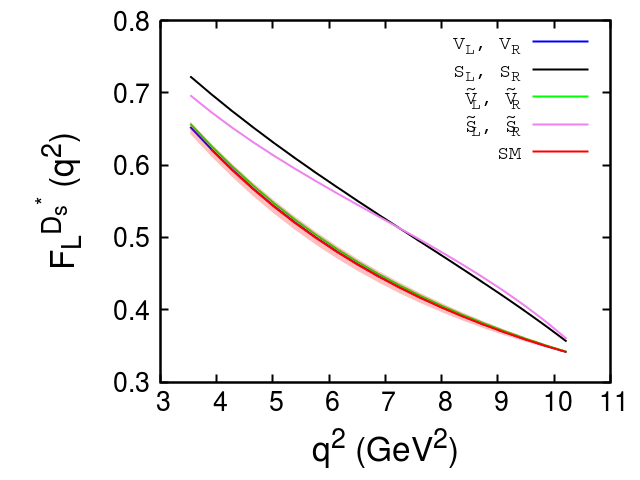}
\caption{$q^2$ dependency of $R_{D_s^*}(q^2)$, $DBR(q^2)$, $A_{FB}^{\tau}(q^2)$, $P_{\tau}^{D_s^*}(q^2)$ 
$A_{FB}^T(q^2)$, $C_F^{\tau}(q^2)$ and $F_{L}^{D_s^*}(q^2)$ for $B_s\rightarrow D_s^* {\tau} \nu$ decay mode in the SM and in the presence of several NP cases. 
The SM central curve and the corresponding error band are shown with red colour. The NP effects due to ($V_L$, $V_R$), ($S_L$, $S_R$),
$(\widetilde{V}_L$, $\widetilde{V}_R$) and ($\widetilde{S}_L$, $\widetilde{S}_R$) are shown with blue, black, green and violet colour
respectively. }
\label{np2plots}
\end{figure}

In Fig.~\ref{np2plots}, we show the $q^2$ dependency of all the observables  
for the $B_s\rightarrow D_s^* {\tau} \nu$ decay mode in the SM as well as in the presence of NP from $2D$ scenario. 
The SM central curve is shown by the red solid line whereas the corresponding 
best fits with ($V_L,V_R$), ($S_L$, $S_R$), ($\widetilde{V}_L$,
$\widetilde{V}_R$) and ($\widetilde{S}_L$ , $\widetilde{S}_R$) NP couplings are shown with blue, black, green and violet lines,
respectively. The observations 
pertaining to the $B_s\rightarrow D_s^*{\tau}\nu$ decay mode are as follows.

\begin{itemize}  
\item In case of $R_{D_s^*}(q^2)$ and ${\rm DBR}(q^2)$, a significant deviation from the SM prediction is observed in each NP scenarios.
It should also be noted that they are clearly distinguishable from the SM prediction.  
 
\item The SM zero crossing in $A^{\tau}_{FB}(q^2)$ is observed at $q^2 \approx 5.2\pm0.2\,\  {\rm GeV^2}$.
We notice that NP couplings ($V_L$, $V_R$) and 
($\widetilde{S}_L$, $\widetilde{S}_R$ ) show similar behavior with the SM prediction.
The zero crossing due to the ($\widetilde{V}_L$, $\widetilde{V}_R$) NP coupling is shifted to a higher value of 
$q^2\approx 6.0\,  {\rm GeV^2}$ which is distinguishable from the SM prediction at more than $4\sigma$ significance. 
Similarly, for the ($\widetilde{S}_L$, $\widetilde{S}_R$) NP coupling, the zero crossing is observed at $q^2\,\approx 3.9\,{\rm GeV^2}$ and 
it is distinguishable from the SM prediction at more than $6\sigma$ significance.
  
\item  In case of $P^{\tau}(q^2)$, we observe significant deviation from the SM prediction in scenarios with 
($\widetilde{V}_L$, $\widetilde{V}_R$), ($S_L$, $S_R$) and ($\widetilde{S}_L$, $\widetilde{S}_R$) NP couplings. 
In case of convexity parameter $C_F^{\tau}(q^2)$, we observe slight deviation from the SM prediction with ($S_L$, $S_R$) and
($\widetilde{S}_L$, $\widetilde{S}_R$) NP couplings in the low $q^2$ region.

\item In case of $A_{FB}^T (q^2)$, a significant deviation from the SM prediction is
observed with ($V_L,V_R$) NP couplings, whereas, the deviation observed with the rest of the NP couplings lies within the SM error band.
Similarly for the longitudinal polarization fraction of $D_s^*$ meson $F_L^{D_s^{\ast}}$, the deviation from the SM prediction is more
pronounced in case of ($S_L$, $S_R$) and ($\widetilde{S}_L$, $\widetilde{S}_R$) NP couplings, whereas, the deviation observed 
with the vector NP couplings lies within the SM error band.
\end{itemize}

\section{Conclusion}

\label{Conclusion}

Experimental measurements of various flavor ratios such as $R_D$, $R_{D^*}$, $R_{J/\psi}$, $P_{\tau}^{D^*}$ and $F_L^{D^*}$ in 
$B\,\to\, D^{(*)}\,\tau\,\nu$ and $B_c\,\to\, J/\psi\,\tau\,\nu$ decays mediated via $b \to c\tau\nu$ quark level transitions differ from 
the SM expectation. If it persists in future experiments, 
it would be a definite signal for beyond the SM physics. In this context, we use a model independent effective theory approach in the
presence of NP to investigate the anomalies present in $b\, \to \,c\tau\,\nu$ quark level transition decays. We consider total eight $1D$
NP scenarios and four $2D$ scenarios for our analysis.
In order to determine the NP scenarios that best explain the data, we perform a combined $\chi^2$ fit by including all the recent 
experimental measurements such as $R_D$, $R_{D^*}$, $R_{J/\psi}$, $P_{\tau}^{D^*}$ and $F_L^{D^*}$ in our analysis. We find the best
estimates of all the observables pertaining to $B\rightarrow D^{(*)}\tau\nu$ and $B_c\,\to \,J/\psi\,\tau\,\nu$ decays. 
Similarly, we also study the $B_s\, \to\, D_s^{*}\,l\,\nu$ decay mode and give predictions of
various physical observable such as $R_{D_s^*}$, $DBR$ $P_{\tau}^{D_s^*}$, $F_{L}^{D_s^*}$, $A_{FB}^{\tau}$, $A_{FB}^T$, $C_F^{\tau}$ 
in the SM and in the presence of NP in $1D$ and $2D$ scenarios.
We also discuss the $q^2$ dependency of each physical observables and study the implication of $b \,\to\, c\,\tau\,\nu$ flavor anomalies within
various $1D$ and $2D$ scenarios. 
Precise measurement of $B_s\,\to D_s^{*}\,l\,\nu$ decay observables in future can, in principle,  provide better understanding of the 
LFUV in $b \to c\tau\nu$ transition decays. Moreover, a precise measurement of the $B_s\rightarrow D_s^{*}l\nu$ branching ratio will allow us 
to determine the not so precise CKM matrix element $|V_{cb}|$.

\end{document}